\documentclass[a4paper,10pt]{article}
\usepackage[dvips]{graphicx}
\usepackage{amssymb,amsmath}
\begin{document}

\title{ Cosmology of F(T)  gravity and k-essence}
\author{Ratbay Myrzakulov\footnote{Email: rmyrzakulov@gmail.com; rmyrzakulov@csufresno.edu} \\ \textit{Eurasian International Center for Theoretical Physics and  Department of General } \\ \textit{ $\&$  Theoretical Physics, Eurasian National University, Astana 010008, Kazakhstan}}

%\date{}

\date{}
\maketitle
\begin{abstract}
This a brief review on $F(T)$ gravity and its relation with
k-essence. Modified teleparallel gravity theory with the torsion
scalar has recently gained a lot of attention as a possible
explanation of dark energy. We perform a thorough  reconstruction
analysis on the so-called $F(T)$ models, where $F(T)$ is some
general function of the torsion term,  and deduce the required
conditions for the equivalence between of $F(T)$ models  with pure
kinetic k-essence models. We present a new class of models of
$F(T)$-gravity and  k-essence.
\end{abstract}
\vspace{2cm}

\sloppy

\tableofcontents
\section{Introduction}
Recent astrophysical data imply that the current expansion of the
universe is accelerating~\cite{Perl}. There exist different
candidates for this acceleration phase. The simplest one is the
introduction of the Cosmological Constant $\Lambda $  in the
framework of General Relativity ($\Lambda$CDM model), namely an
exotic form of energy (the dark energy) whose Equation of State
(EoS) parameter $w$ is equal to minus one and dynamically remains
near this value, but in principle quintessence/phantom-fluid
description is not excluded. Despite the fact that the $\Lambda$CDM
is a good candidate to describe our universe, the finite but very
small value of the $\Lambda $  causes some well-known problems,
 like the difference between the order of $\Lambda $  predicted by quantum field theory  which is so called by fine-tuning, another problem is the time where such acceleration happen which is the coincidence problem.
Further, the origin of dark energy is an unsolved question. Also,
the existence of an early accelerated epoch, namely the inflation,
introduces a new problem to the standard cosmology, and various
proposals to construct acceptable inflationary model which exist
like the scalar, spinor $SU(2)$, (non-)abelian vector theory
($SU(2)$) $U(1)$ and so on.

Another alternative approach to the dark energy puzzle is the
modified gravity theories. A typical  modified gravity is a
generalization of Einstein's gravity, where some combination of
curvature invariants is added into the classical Hilbert-Einstein
action of General Relativity. This modification may lead to an
accelerated era without invoking the dark energy. The simplest
theory of modified gravity is the $F(R)$ one, where the modification
is given by a function of the Ricci scalar only. Another popular
modification is given by the string-inspired Gauss-Bonnet modified
theories, where a modification via the topological invariant four
dimensional Gauss Bonnet  $G$ appears (see the recent reviews
\cite{N1}-\cite{CST}). Also it can be represented by the $f(R,T)$
models where $T$ is the trace of the energy-momentum tensor
\cite{frt1}-\cite{frt3}. The field equations of these theories are
much more complicated with respect to the case of General
Relativity, since they are 4th order differential equations and it
is so difficult to obtain the  exact  solutions.

Recently a new type of gravity model, the $F(T)$-gravity, has been
proposed. Its  field equations are  2nd order \cite{BF}-\cite{L3}.
These models are based on the "teleparallel" equivalent of General
Relativity (TEGR) \cite{Eins}-\cite{FF2}, which, instead of using
the curvature defined via the Levi-Civita connection, uses the
Weitzenb$\ddot{o}$ck connection that has no curvature but only
torsion (see Refs. \cite{FF1}-\cite{FF2} for applications to
inflation). The fact that the field equations of $F(T)$ gravity are
2nd order makes these theories simpler than the  ones where
modification is via curvature invariants and it is of extreme
interest a deeper investigation on this kind of models (see Refs.
\cite{M6}-\cite{epjc4} for recent developments).

In this  paper we give  a brief review on $F(T)$ gravity and its relation with k-essence. We study  some $F(T)$ -models and models of k-essence.
In the  following sections 2 and 3 we present some  basic facts on  $F(T)$ gravity. In the section 4, we study  some models of $F(T)$ gravity for the FRW spacetime. Noether symmetry in $F(T)$ gravity was considered in the section 5. In Sec. 5, we consider the torsion-scalar model.   We investigate  k-essence and its models in Sec. 7 and in Sec. 8.  Sec. 9 is devoted to the study of the relation between $F(T)$ gravity and k-essence and in Sec. 10 we present some generalizations of $F(T)$ gravity.  In the last section we give conclusions and general remarks.

\section{General aspects of $F(T)$ gravity}
The action of   $F(T)$ - gravity   reads \cite{BF, L3, M6}
\begin {equation}
S=\int e \mathcal{L}d^{4}x,
\end{equation}
where
\begin {equation}
\mathcal{L}=\frac{1}{2\kappa^{2}}F(T)+\mathcal{L}_{m}.
\end{equation}Here $T$ is the torsion scalar, $e=\det{(e^{i}_{\mu})}=\sqrt{-g}$ and $\mathcal{L}_{m}$ is the matter Lagrangian. Here $e^{i}_{\mu}$ are the components of the vierbein vector field $\textbf{e}_{A}$ in the coordinate basis $\textbf{e}_{A}\equiv e^{\mu}_{A}\partial_{\mu}$. Note that in the teleparallel gravity, the dynamical variable is the vierbein field $\textbf{e}_{A}(x^{\mu})$. To derive the equations of motion we consider the metric
\begin{equation}
ds^2=g_{\mu\nu}dx^{\mu}dx^{\nu}=\eta_{ab}\theta^{a}\theta^{b}\label{ele},
\end{equation}
where
\begin{equation}
\theta^{a}=e^{a}_{\;\;\mu}dx^{\mu}\;,\;dx^{\mu}=e_{a}^{\;\;\mu}\theta^{a}\label{the},
\end{equation}
$g_{\mu\nu}$ being the metric of space-time, $\eta_{ab}$ the Minkowski's metric, $\theta^{a}$ the tetrads and $e^{a}_{\;\;\mu}$ and their inverses $e_{a}^{\;\;\mu}$ the tetrads basis. We note that the tetrad basis satisfy the relations
 \begin{equation}
 e^{a}_{\;\;\mu}e_{a}^{\;\;\nu}=\delta^{\nu}_{\mu}, \quad e^{a}_{\;\;\mu}e_{b}^{\;\;\mu}=\delta^{a}_{b}.
\end{equation} The root of the metric determinant is given by
 \begin{equation}
 e=\sqrt{-g}=det[e^{a}_{\;\;\mu}].
\end{equation}  The standard Weitzenbok's connection reads
\begin{equation}
\Gamma^{\alpha}_{\mu\nu}=e_{i}^{\;\;\alpha}\partial_{\nu}e^{i}_{\;\;\mu}=-e^{i}_{\;\;\mu}\partial_{\nu}e_{i}^{\;\;\alpha}\label{co}.
\end{equation}
As a result, the covariant derivative, denoted by $D_\mu$, satisfies the equation
\begin{equation}
D_\mu e^i_\nu=\partial_{\mu}e^{i}_{\nu}-\Gamma^{\lambda}_{\nu\mu}e^{i}_{\lambda}=0.
\end{equation}
Then  the components of the torsion and the contorsion are given by
\begin{eqnarray}
T^{\alpha}_{\;\;\mu\nu}&=&\Gamma^{\alpha}_{\nu\mu}-\Gamma^{\alpha}_{\mu\nu}=e_{i}^{\;\;\alpha}\left(\partial_{\mu} e^{i}_{\;\;\nu}-\partial_{\nu} e^{i}_{\;\;\mu}\right)\label{tor}\;,\\
K^{\mu\nu}_{\;\;\;\;\alpha}&=&-\frac{1}{2}\left(T^{\mu\nu}_{\;\;\;\;\alpha}-T^{\nu\mu}_{\;\;\;\;\alpha}-T_{\alpha}^{\;\;\mu\nu}\right)\label{cont}\; .
\end{eqnarray}
Now we define another tensor from the components of torsion and the contorsion as
\begin{equation}
S_{\alpha}^{\;\;\mu\nu}=\frac{1}{2}\left( K_{\;\;\;\;\alpha}^{\mu\nu}+\delta^{\mu}_{\alpha}T^{\beta\nu}_{\;\;\;\;\beta}-\delta^{\nu}_{\alpha}T^{\beta\mu}_{\;\;\;\;\beta}\right)\label{s}\;.
\end{equation}
Finally, we  define  the torsion scalar as usual
\begin{equation}
T=T^{\alpha}_{\;\;\mu\nu}S_{\alpha}^{\;\;\mu\nu}\label{t1},
\end{equation}
Let us derive the equations of motion from the Euler-Lagrange equations. In order to use these equations we first write the quantities
\begin{equation}
\frac{\partial L}{\partial e^{a}_{\;\;\mu}}=F(T)ee_{a}^{\;\;\mu}+eF_{T}(T)4e_{a}^{\;\;\alpha}T^{\sigma}_{\;\;\nu\alpha}S_{\sigma}^{\;\;\mu\nu}+\frac{\partial{L}_{m}}{\partial e^{a}_{\;\;\mu}},\label{1}
\end{equation}
and
\begin{equation}
\partial_{\alpha}\left[\frac{\partial {L}}{\partial (\partial_{\alpha}e^{a}_{\;\;\mu})}\right]=-4F_{T}(T)\partial_{\alpha}\left(ee_{a}^{\;\;\sigma}S_{\sigma}^{\;\;\mu\nu}\right)-4ee_{a}^{\;\;\sigma}S_{\sigma}^{\;\;\mu\alpha}\partial_{\alpha}T\,F_{TT}(T)+\partial_{\alpha}\left[\frac{\partial {L}_{m}}{\partial (\partial_{\alpha}e^{a}_{\;\;\mu})}\right]\label{2},
\end{equation}
where $F_{T}(T)=dF(T)/dT$ and $F_{TT}(T)=d^2F(T)/dT^2$. Now we use the  Euler-Lagrange equation
\begin{eqnarray}\label{EL}
\frac{\partial\mathcal{L}}{\partial e^{a}_{\;\;\mu}}-\partial_{\alpha}\left[\frac{\partial \mathcal{L}}{\partial (\partial_{\alpha}e^{a}_{\;\;\mu})}\right]=0\;.
\end{eqnarray}
Substituting the expressions \eqref{1} and \eqref{2} into the later
equation we get the equations of motion of the $F(T)$ gravity (after
mulltiplying by $e^{-1}e^{a}_{\;\;\beta}/4$)
\begin{eqnarray}
S_{\beta}^{\;\;\mu\alpha}\partial_{\alpha}T\,F_{TT}(T)+\left[e^{-1}e^{a}_{\;\;\beta}\partial_{\alpha}\left(ee_{a}^{\;\;\sigma}S_{\sigma}^{\;\;\mu\alpha}\right)+T^{\sigma}_{\;\;\nu\beta}S_{\sigma}^{\;\;\mu\nu}\right]F_{T}(T)+\frac{1}{4}\delta^{\mu}_{\beta}F(T)=4\pi \mathcal{T}^{\mu}_{\beta}\label{em}\;,
\end{eqnarray}
where
\begin{eqnarray}
\mathcal{T}^{\mu}_{\beta}=-\frac{e^{-1}e^{a}_{\;\;\beta}}{16\pi}\left\{ \frac{\partial \mathcal{L}_{Matter}}{\partial e^{a}_{\;\;\mu}}-\partial_{\alpha}\left[\frac{\partial \mathcal{L}_{Matter}}{\partial (\partial_{\alpha}e^{a}_{\;\;\mu})}\right]\right\}
\end{eqnarray}
is the gravitational energy momentum tensor. Of course, if we consider the TG case that is  $F(T)=T$ then  the gravitional equations reduce to
\begin{equation}
T-2e^{-1}\partial_{\sigma}(eT_{\rho}{}^{\rho\sigma})=\kappa^2\mathcal{T}^{\mu}_{\mu},
\end{equation}
which shows an equivalence between GR and TG since
\begin{equation}
-R=T-2e^{-1}\partial_{\sigma}(eT_{\rho}{}^{\rho\sigma}).
\end{equation}
%%%%%%%%%%%%%%%%%%%%%%%%%%%%%%%%%%%%%%%%%%%%%%%%%%%%%%%%%%%%%%%%%%%%%%%%%%%%%%%%%%%%%%%%%%%%%%%%%%%%%%%%

\section{The FRW space-time}

We will assume a flat homogeneous and isotropic FRW universe with  the metric
\begin{equation}
ds^{2}=-dt^{2}+a(t)^{2}\sum^{3}_{i=1}(dx^{i})^{2},
\end{equation}
where $t$ is  cosmic time and $a(t)$ is the scale factor. Then the modified Friedmann equations and the continuity equation read (see, e.g. \cite{BF}, \cite{L3}, \cite{M6})
\begin{eqnarray}
    -2TF_{T}+F&=&2\kappa^2 \rho_m, \\
    -8\dot{H}TF_{TT}+(2T-4\dot{H})F_{T}-F&=&2\kappa^2p_m, \\
    \dot{\rho}_m+3H(\rho_m+p_m)&=&0.
\end{eqnarray}
This set  can be rewritten as
\begin{eqnarray}
    -T-2Tf_{T}+f&=&2\kappa^2 \rho_m, \label{00}\\
        -8\dot{H}Tf_{TT}+(2T-4\dot{H})(1+f_{T})-T-f&=&2\kappa^2p_m, \\
    \dot{\rho}_m+3H(\rho_m+p_m)&=&0, \label{11}
\end{eqnarray}
if we consider the following equivalent form of the action
\begin {equation}
S=\int d^{4}xe[\frac{1}{2\kappa^{2}}(T+f(T))+\mathcal{L}_{m}],
\end{equation} where $f=F-T.$  Some properties of $F(T)$ - gravity  were studied in \cite{L3}-\cite{M7}.
The field equations (\ref{00})-(\ref{11}) are equivalent to
\begin{eqnarray}
    \hat{M}_1F&=&2\kappa^2 \rho_m, \\
\hat{M}_2F&=&-\hat{M}_{3}\hat{M}_{1}F=2\kappa^2p_m, \\
\hat{M}_3\rho_{m}&=&-p_{m},
\end{eqnarray}
where
\begin{eqnarray}
\hat{M}_1&=&    -2T\partial_{T}+1, \\
\hat{M}_2&=&    -8\dot{H}T\partial^2_{TT}+(2T-4\dot{H})\partial_{T}-1=(4\dot{H}\partial_{T}-1)\hat{M}_{1}=-(\frac{1}{3H}\partial_{t}+1)\hat{M}_{1}
=-\hat{M}_{3}\hat{M}_{1},\nonumber\\ \\
\hat{M}_3&=&    \frac{1}{3H}\partial_{t}+1.
\end{eqnarray}
By using these equations we may construct high hierarchy of $F(T)$ gravity. For the case $\rho_m=p_m=0$ such hierarchy is written as
\begin{equation}
\hat{M}^n_1F_n=0,
\end{equation}
where $F_1=F$ and (for $n=1,2,3$)
\begin{eqnarray}
-2TF_{1T}+F_1&=&0, \\
4T^2F_{2TT}+F_2&=&0, \\
-8T^3F_{3TTT}-12T^2F_{3TT}-2TF_{3T}+F_3&=&0,
\end{eqnarray}
and so on. From the system (2.16)-(2.18) one has that any solution of the equation (2.16) automatically solves the equations (2.17)-(2.18). It means
 that by solving the equation (2.16), we have also a solution for the equations (2.17) and (2.18).  Finally we introduce the effective EoS parameter
 \begin{equation}
w_{eff}=-1-3^{-1}H^{-1}[\ln{(\hat{M}_1F)}]_{t}=-1-3^{-1}[\ln{(\hat{M}_1F)}]_{N}.
\end{equation}

\section{Specific models of $F(T)$ gravity in FRW universe}

Some explicit models of $F(T)$ gravity have recently appeared in the literature (see, e.g. \cite{BF}, \cite{L3}, \cite{M6}, \cite{M3}, \cite{Yang4}, \cite{M1}, \cite{WuYu3}, \cite{epjc1}). Here, we would like to present some new  models of modified teleparallel gravity.

 \subsection{Example 1: The M$_{13}$ - model}

 Let us consider the  M$_{13}$ - model. Its   Lagrangian is
 \begin {equation}
F(T)=\sum_{j=-m}^{n}\nu_j(t)T^{j}=\nu_{-m}(t)T^{-m}+ ... +\nu_{-1}(t)T^{-1}+\nu_0(t)+\nu_{1}(t)T + ... + \nu_n(t)T^{n}.
\end{equation}
We consider the particular case where $m=n=1$ and $\nu_j=\textbf{consts}$. Thus,
 \begin {equation}
F=\nu_{-1}T^{-1}+\nu_{0}+\nu_{1}T, \quad
F_{T}=-\nu_{-1}T^{-2}+\nu_{1}, \quad F_{TT}=2\nu_{-1}T^{-3}.
\end{equation}
By substituting these expressions into (2.9)-(2.10) we obtain
\begin{align}
3\kappa^{-2}H^2=\rho_{eff}+ \rho_m,
\end{align}
\begin{align}
-\kappa^{-2}(2\dot{H}+3H^2)=p_{eff}+p_m,
\end{align}
where
\begin{align}
\rho_{eff}=\kappa^{-2}[3H^2-1.5\nu_{-1}T^{-1}+0.5\nu_1 T-0.5\nu_0],
\end{align}
\begin{align}
p_{eff}=\kappa^{-2}[6\nu_{-1}\dot{H}T^{-2}+1.5\nu_{-1} T^{-1}-0.5\nu_1T+0.5\nu_0+2(\nu_1-1)\dot{H}-3H^2].
\end{align}
The effective EoS parameter is given by
\begin{align}
w_{eff}=\frac{p_{eff}}{\rho_{eff}}=\frac{6\nu_{-1}\dot{H}T^{-2}+1.5\nu_{-1} T^{-1}-0.5\nu_1T+0.5\nu_0+2(\nu_1-1)\dot{H}-3H^2}{3H^2-1.5\nu_{-1}T^{-1}+0.5\nu_1 T-0.5\nu_0}.
\end{align}
Let us set $\nu_1=1$. Thus,
\begin{align}
\rho_{eff}=\kappa^{-2}[-1.5\nu_{-1}T^{-1}-0.5\nu_0], \quad
p_{eff}=\kappa^{-2}[6\nu_{-1}\dot{H}T^{-2}+1.5\nu_{-1} T^{-1}+0.5\nu_0]
\end{align}
and
\begin{align}
w_{eff}=\frac{p_{eff}}{\rho_{eff}}=\frac{6\nu_{-1}\dot{H}T^{-2}+1.5\nu_{-1} T^{-1}+0.5\nu_0}{-1.5\nu_{-1}T^{-1}-0.5\nu_0}=-1-\frac{6\nu_{-1}\dot{H}T^{-2}}{1.5\nu_{-1}T^{-1}+0.5\nu_0}\,.
\end{align}

\subsection{Example 2: The M$_{21}$ - model}
 Our next example is the M$_{21}$ - model
  \begin{align}
F=T+\alpha T^{\delta}\ln{T}.
\end{align}
Now
  \begin {equation}
F_{T}=1+\alpha\delta T^{\delta-1}\ln{T}+\alpha T^{\delta-1}, \quad
F_{TT}=\alpha\delta(\delta-1) T^{\delta-2}\ln{T}+\alpha(2\delta-1) T^{\delta-2}.
\end{equation}
As a consequence, Eqs.(2.9)-(2.10) take the form
 \begin{align}
    -T-2\alpha T^{\delta}-\alpha(2\delta-1) T^{\delta}\ln{T}=2\kappa^2 \rho_m,
\end{align}
\begin{align}
\alpha(2\delta-1)(T-4\delta\dot{H})T^{\delta-1}\ln{T}+T-4\dot{H}+2\alpha T^{\delta}-4\alpha\dot{H}(4\delta-1)T^{\delta-1}=2\kappa^2p_m.
\end{align}
One has
\begin{align}
\rho_{eff}=0.5\kappa^{-2}[2\alpha T^{\delta}+\alpha(2\delta-1) T^{\delta}\ln{T}],
\end{align}
\begin{align}
p_{eff}=-0.5\kappa^{-2}\alpha T^{\delta-1}[(2\delta-1)(T-4\delta\dot{H})\ln{T}+2T-4(4\delta-1)\dot{H}].
\end{align}
The special case $\delta=1/2$ deserves a separate consideration. In
this case the above equations take a simpler form
\begin{align}
    -T-2\alpha T^{0.5}=2\kappa^2 \rho_m, \quad
    T-4\dot{H}+2\alpha T^{0.5}-4\alpha\dot{H}T^{-0.5}=2\kappa^2p_m.
\end{align}
For the effective energy density and pressure we get
\begin{align}
\rho_{eff}=\kappa^{-2}\alpha T^{0.5}, \quad p_{eff}=-\kappa^{-2}\alpha T^{-0.5}(T-2\dot{H}).
\end{align}

\subsection{Example 3: The M$_{22}$ - model}
 Now we consider  the M$_{22}$ - model
  \begin{align}
F=T+f(y), \quad y=\tanh[T].
\end{align}
Thus
  \begin {equation}
F_{T}=1+f_y(1-y^2),\quad
F_{TT}=f_{yy}(1-y^2)^2-2y(1-y^2)f_y
\end{equation}
so that  Eqs.(2.9)-(2.10) take the form
 \begin{align}
    -T-2(1-y^2)Tf_y+f=2\kappa^2 \rho_m,
\end{align}
\begin{align}
    T-4\dot{H}-8(1-y^2)^2T\dot{H}f_{yy}+(16y\dot{H}T+2T-4\dot{H})(1-y^2)f_{y}-f=2\kappa^2p_m.
\end{align}
We have
\begin{align}
\rho_{eff}=0.5\kappa^{-2}[2(1-y^2)Tf_y-f],
\end{align}
\begin{align}
p_{eff}=0.5\kappa^{-2}[8(1-y^2)^2T\dot{H}f_{yy}-(16y\dot{H}T+2T-4\dot{H})(1-y^2)f_{y}+f].
\end{align}
The EoS parameter  reads
$$
w_{eff}=\frac{8(1-y^2)^2T\dot{H}f_{yy}-(16y\dot{H}T+2T-4\dot{H})(1-y^2)f_{y}+f}{2(1-y^2)Tf_y-f}=
$$
\begin{align}
=-1+\frac{8(1-y^2)^2T\dot{H}f_{yy}-(16y\dot{H}T-4\dot{H})(1-y^2)f_{y}+f}{2(1-y^2)Tf_y-f}.
\end{align}

\subsection{Example 4: The M$_{25}$ - model}
 In this subsection  we will consider  the M$_{25}$ - model
  \begin{equation}
F=\sum_{-m}^{n}\nu_{j}(t)\xi^{j},
\end{equation}
where $\xi=\ln{T}$. We take the case   $m=n=1$ and $\nu_j=\textbf{consts}$, namely
  \begin {equation}
F=\nu_{-1}\xi^{-1}+\nu_0+\nu_1\xi.
\end{equation}
Thus
  \begin {equation}
F_{\xi}=-\nu_{-1}\xi^{-2}+\nu_1,\quad
F_{\xi\xi}=2\nu_{-1}\xi^{-3}
\end{equation}
and
  \begin {equation}
F_{T}=(-\nu_{-1}\xi^{-2}+\nu_1)e^{-\xi},\quad
F_{TT}=(2\nu_{-1}\xi^{-3}+\nu_{-1}\xi^{-2}-\nu_1)e^{-2\xi}.
\end{equation}
In this case, Eqs.(2.9)-(2.10) lead to
 \begin{equation}
    2\nu_{-1}\xi^{-2}+\nu_{-1}\xi^{-1}+\nu_0-2\nu_1+\nu_1\xi=2\kappa^2 \rho_m,
\end{equation}
\begin{equation}
    -4\dot{H}(4\nu_{-1}\xi^{-3}+\nu_{-1}\xi^{-2}-\nu_1)e^{-\xi}-2\nu_{-1}\xi^{-2}-\nu_{-1}\xi^{-1}+2\nu_1-\nu_0-\nu_1\xi=2\kappa^2p_m.
\end{equation}

    \section{Noether symmetry in $F(T)$ gravity}

In this section we want to present a brief review on Noether symmetry in $F(T)$ gravity following to the  paper \cite{Hao}. Generally speaking, Noether
 symmetry is a power method  to select models motivated at a
 fundamental level. It also allows to construct  the exact solution of the model. We start from the point-like Lagrangian of $F(T)$ gravity:
\begin{equation}\label{eq15}
 {\cal L}(a,\dot{a},T,\dot{T})=a^3\left(f-f_T T\right)-
 6f_T a\dot{a}^2-\rho_{m0}.
 \end{equation}
We now use  the Euler-Lagrange
 equation:
 \begin{equation}\label{eq16}
 \frac{d}{dt}\left(\frac{\partial \cal L}{\partial\dot{q}_i}
 \right)-\frac{\partial \cal L}{\partial q_i}=0\,,
\end{equation}
 where $q_i$ are the generalized coordinates of
 the phase space and  $q_i=a$
 and $T$. Then we have
\begin{eqnarray}
 a^3 f_{TT}\left(T+
 6\frac{\dot{a}^2}{a^2}\right)&=&0\,,\label{eq17}\\
 f-f_T T+2f_T H^2+
 4\left(f_T\frac{\ddot{a}}{a}+Hf_{TT}\dot{T}\right)&=&0\,.\label{eq18}
\end{eqnarray}
 Hence as $f_{TT}\not=0$ we obtain
\begin{equation}\label{eq19}
 T=-6\frac{\dot{a}^2}{a^2}=-6H^2
\end{equation}
 that is  the Euler constraint of the dynamics. Next using
 $\ddot{a}/a=H^2+\dot{H}$, we obtain
\begin{equation}\label{eq20}
 48H^2 f_{TT}\dot{H}-4f_T\left(3H^2+\dot{H}\right)-f=0\,,
\end{equation}
 i.e., the modified second Friedmann  equation. Now let us consider  the Hamiltonian corresponding
 to Lagrangian $\cal L$ \cite{Hao}:
\begin{equation}\label{eq21}
 {\cal H}=\sum\limits_i\frac{\partial \cal L}
 {\partial\dot{q}_i}\dot{q}_i-{\cal L}
\end{equation}
 so that
\begin{equation}\label{eq22}
 {\cal H}(a,\dot{a},T,\dot{T})=a^3\left(-6f_T
 \frac{\dot{a}^2}{a^2}-f+f_T T+\frac{\rho_{m0}}{a^3}\right).
\end{equation}
 Assuming that  the total energy ${\cal H}=0$ (Hamiltonian
 constraint) and from  Eq.~(\ref{eq19}),
 we get
\begin{equation}\label{eq23}
 12H^2 f_T+f=\frac{\rho_{m0}}{a^3}
\end{equation}
that is nothing but the first Friedmann equation.

Now we want to present the Noether symmetry for $F(T)$ gravity in the FRW metric case. To do it, we introduce  the generator of Noether
 symmetry as \cite{Hao}\begin{equation}\label{eq24}
 {\bf X}=\alpha\frac{\partial}{\partial a}+
 \beta\frac{\partial}{\partial T}+
 \dot{\alpha}\frac{\partial}{\partial\dot{a}}+
 \dot{\beta}\frac{\partial}{\partial\dot{T}}\,,
\end{equation}
 where $\alpha=\alpha(a,T)$ and $\beta=\beta(a,T)$. As is well known, Noether
 symmetry exists if the equation
\begin{equation}\label{eq25}
 L_{\bf X}{\cal L}={\bf X}{\cal L}=
 \alpha\frac{\partial \cal L}{\partial a}+
 \beta\frac{\partial \cal L}{\partial T}+
 \dot{\alpha}\frac{\partial \cal L}{\partial\dot{a}}+
 \dot{\beta}\frac{\partial \cal L}{\partial\dot{T}}=0
\end{equation}
 has solution. Here $L_{\bf X}{\cal L}$ is the Lie derivative
 of the Lagrangian $\cal L$ with respect to the vector $\bf X$.
 The corresponding Noether charge reads as \cite{Hao}
\begin{equation}\label{eq26}
 Q_0=\sum\limits_i\alpha_i\frac{\partial \cal L}{\partial\dot{q}_i}
 =\alpha\frac{\partial \cal L}{\partial\dot{a}}+
 \beta\frac{\partial \cal L}{\partial\dot{T}}=const.
\end{equation}
 From  Eq.~(\ref{eq25}) and using the relations $\dot{\alpha}=
 (\partial\alpha/\partial a)\,\dot{a}
 +(\partial\alpha/\partial T)\,\dot{T}$, $\dot{\beta}=
 (\partial\beta/\partial a)\,\dot{a}+
 (\partial\beta/\partial T)\,\dot{T}$, we come to the equation \cite{Hao}
\begin{equation}\label{eq27}
 3\alpha a^2\left(f-f_T T\right)-\beta a^3 f_{TT}T
 -6\dot{a}^2\left(\alpha f_T+\beta a f_{TT}+
 2a f_T\frac{\partial\alpha}{\partial a}\right)-
 12a\dot{a}\dot{T}\frac{\partial\alpha}{\partial T}=0\,.
\end{equation}
 Now we  impose that the coefficients of $\dot{a}^2$,
 $\dot{T}^2$ and $\dot{a}\dot{T}$ in Eq.~(\ref{eq27}) to be
 zero. Then we get \cite{Hao}
 \begin{eqnarray}
 a\frac{\partial\alpha}{\partial T}&=&0\,,\label{eq28}\\
 \alpha f_T+\beta a f_{TT}+
 2a f_T\frac{\partial\alpha}{\partial a}&=&0\,,\label{eq29}\\
 3\alpha a^2\left(f-f_T T\right)-
 \beta a^3 f_{TT}T&=&0\,.\label{eq30}
\end{eqnarray}
 As is known, the constraint~(\ref{eq30}) is sometimes
 called Noether condition. The corresponding Noether charge looks like \cite{Hao}
\begin{equation}\label{eq31}
 Q_0=-12\alpha f_T a\dot{a}=const.
\end{equation}
 From  Eq.~(\ref{eq28})  it follows that  $\alpha=\alpha(a)$.
 On the other hand,  Eq.~(\ref{eq30}) gives us
\begin{equation}\label{eq32}
 \beta a f_{TT}T=3\alpha\left(f-f_T T\right).
\end{equation}
 Hence  we have
\begin{equation}\label{eq33}
 f_T T\left(2a\frac{d\alpha}{da}-2\alpha\right)+3\alpha f=0,
\end{equation}
 which we  recast  as
\begin{equation}\label{eq34}
 1-\frac{a}{\alpha}\frac{d\alpha}{da}=\frac{3f}{2f_T T}\,.
\end{equation}
 The last  equations we split into two equations write as \cite{Hao}
\begin{eqnarray}
 nf-f_T T&=&0,\label{eq35}\\
 1-\frac{a}{\alpha}\frac{d\alpha}{da}-\frac{3}{2n}&=&0.\label{eq36}
 \end{eqnarray}
 These equations have the solutions  \cite{Hao}
\begin{eqnarray}
 f(T)&=&\mu T^n\,,\label{eq37}\\
 \alpha(a)&=&\alpha_0\,a^{1-3/(2n)}\,,\label{eq38}
 \end{eqnarray}
 where $\mu$ and $\alpha_0$ are real  constants. So from  Eq.~(\ref{eq32}), we get
\begin{equation}\label{eq39}
 \beta(a,T)
 =-\frac{3\alpha_0}{n}\,a^{-3/(2n)}\,T\,.
\end{equation}
 Finally we can conclude that the explicit non-zero solutions of
 $f(T)$, $\alpha$ and $\beta$  exists that means  Noether symmetry
 exists. Note that Noether symmetry allows us to construct  the exact solution of
 $a(t)$ for the given  $f(T)$ model. For example, from
 Eq.~(\ref{eq31}) follows  \cite{Hao}
\begin{equation}\label{eq40}
 a^{c_1}\,\dot{a}=c_2\,,
\end{equation}
 where
\begin{equation}\label{eq41}
 c_1=\frac{3}{2n}-1\,,~~~~~~~
 c_2=\left[\frac{Q_0}{-12\alpha_0\mu n(-6)^{n-1}}
 \right]^{1/(2n-1)}\,.
\end{equation}
 Its  solution reads as
\begin{equation}\label{eq42}
 a(t)=-(1+c_1)(c_3-c_2 t)^{1/(1+c_1)}
 =(-1)^{1+2n/3}\cdot\frac{3}{2n}\,(c_2 t-c_3)^{2n/3}\,,
\end{equation}
 where $c_3=\textbf{conts}$. This solution  describes the accelerated  expansion of the universe as
\begin{equation}\label{eq43}
 a(t)\sim t^{2n/3}\,,
\end{equation}
 where its prefactor
 $(-1)^{1+2n/3}\cdot\frac{3}{2n}\,c_2^{2n/3}$ is not
 important. As is well-known, in order to get the expanding
 universe, the constraint  $n>0$ is required \cite{Hao}.

\section{The torsion-scalar model}

In this section we would like to study the $F(T)$ gravity in the presence of matter whose lagrangian is
\begin{align}
L_m=\frac{1}{2}\epsilon\dot{\phi}^2-V(\phi),
\end{align}
where $\phi$ is a scalar field and $V(\phi)$ the potential depending on $\phi$. The equations of motion assume the form
\begin{align}
    -2TF_{T}+F=2\kappa^2 [\frac{1}{2}\epsilon\dot{\phi}^2+V],
\end{align}
\begin{align}
    -8\dot{H}TF_{TT}+(2T-4\dot{H})F_{T}-F=2\kappa^2[\frac{1}{2}\epsilon\dot{\phi}^2-V],
\end{align}
\begin{align}
    \ddot{\phi}+3H\dot{\phi}+\epsilon\frac{\partial V}{\partial\phi}=0,
\end{align}
where $\epsilon=1$ for the usual case and $\epsilon=-1$ for the phantom case. From this system we get
\begin{align}
\epsilon\dot{\phi}^2=-8\dot{H}TF_{TT}-4\dot{H}F_{T}, \quad V=4\dot{H}TF_{TT}-2(T-\dot{H})F_{T}+F,
\end{align}
where dot denotes the derivative with respect to the time. If we compare these equations with (2.14)-(2.16) we have
\begin{align}
    \rho=\frac{1}{2}\epsilon\dot{\phi}^2+V, \quad p=\frac{1}{2}\epsilon\dot{\phi}^2-V.
\end{align}
For simplicity we restrict ourself to the case   $F=\alpha T+\beta T^{0.5}$. Thus,
\begin{align}
\epsilon\dot{\phi}^2=-4\alpha\dot{H}, \quad V=-\alpha T+2\alpha\dot{H}.
\end{align}
and
\begin{align}
w=-1+4\frac{\dot{H}}{T}.
\end{align}
Let us consider some examples.
%\section{Reconstruction for a given $a(t)$}

\subsection{Example 1: $a=\delta \sinh^{m}[\mu t]$}
In our first example we consider the following form for the scale factor
\begin{align}a=\delta \sinh^{m}[\mu t].
\end{align}
As a consequence
\begin{align}
H=\mu m\coth[\mu t], \quad \dot{H}=-\frac{\mu^2m}{\sinh^2[\mu t]},\quad \dot{\phi}^2=\frac{4\alpha\mu^2m}{\epsilon \sinh^2[\mu t]}.
\end{align}
So we obtain
\begin{align}
\phi=\phi_0\pm2\sqrt{\frac{\alpha\mu^2m}{\epsilon}}\log[\tanh[\frac{\mu t}{2}]],\quad V=6\alpha m^2 \mu^2\coth^2[\mu t]-\frac{2\alpha\mu^2m}{ \sinh^2[\mu t]}\,,
\end{align}
and the potential takes the form ( $\tanh[\frac{\mu t}{2}]=e^{\pm\frac{\phi-\phi_0}{2\sqrt{\alpha\mu^2 m \epsilon^{-1}}}}$)
\begin{align}
V=3\alpha m^2 \mu^2[\frac{1+e^{\pm\frac{\phi-\phi_0}{\sqrt{\alpha\mu^2 m \epsilon^{-1}}}}}{e^{\pm\frac{\phi-\phi_0}{2\sqrt{\alpha\mu^2 m \epsilon^{-1}}}}}]-
\frac{\alpha\mu^2m(1-e^{\pm\frac{\phi-\phi_0}{\sqrt{\alpha\mu^2 m \epsilon^{-1}}}})^2}{2e^{\pm\frac{\phi-\phi_0}{\sqrt{\alpha\mu^2 m \epsilon^{-1}}}}}.
\end{align}

\subsection{Example 2: $a=a_0e^{\beta t^m}$}

Let  us consider the case $a=a_0e^{\frac{\delta}{m+1} t^{m+1}}$. Thus,  $H=\delta t^m$ and we have
\begin{align}
t=[\frac{(\phi-\phi_0)(m+1)}{\pm 4\sqrt{-\alpha m\delta\epsilon^{-1}}}]^{\frac{2}{m+1}},\quad \epsilon\dot{\phi}^2=-4\alpha m\delta t^{m-1},
\end{align}
such that
\begin{align}
\phi=\phi_0\pm\frac{4\sqrt{-\alpha m\delta\epsilon^{-1}}}{m+1}t^{\frac{m+1}{2}}, \quad V=6\alpha\delta^2 t^{2m}+2\alpha m\delta t^{m-1}.
\end{align}
We finally get
\begin{align}
V=6\alpha\delta^2[\frac{(\phi-\phi_0)(m+1)}{\pm4\sqrt{-\alpha m\delta\epsilon^{-1}}}]^{\frac{4m}{m+1}}+2\alpha m \delta [\frac{(\phi-\phi_0)(m+1)}{\pm 4\sqrt{-\alpha m\delta}}]^{\frac{2(m-1)}{m+1}}.
\end{align}

\subsection{Example 3: $a=a_0t^n$}
The next example is given by
\begin{align}
a=a_0t^n\,,
\end{align}
for which
\begin{align}
H=\frac{n}{t}, \quad \dot{H}=-\frac{n}{t^2},\quad \epsilon\dot{\phi}^2=\frac{4\alpha n}{t^2}, \quad \phi-\phi_0=\pm 2\sqrt{\alpha n \epsilon^{-1}}\ln[t],\quad t=e^{\pm\frac{\phi-\phi_0}{2\sqrt{\alpha n \epsilon^{-1}}}}
\end{align}
and
\begin{align}
V=2\alpha n(3n-1)t^{-2}.
\end{align}
The potential assumes the final form
\begin{align}
V=2\alpha n(3n-1)e^{\mp\frac{\phi-\phi_0}{\sqrt{\alpha n\epsilon^{-1}}}}.
\end{align}

\section{The k-essence}

 The action of k-essence reads \cite{3}-\cite{32}
\begin {equation}
S=\int d^{4}x\sqrt{-g}[\frac{1}{2\kappa^{2}}R+K(X,\phi)+L_{m}].
\end{equation}
The corresponding (closed) set of equations for FRW metric (2.8) is
\begin{equation}
3\kappa^{-2}H^2=2XK_{X}-K+\rho_m,
\end{equation}
\begin{equation}
    -\kappa^{-2}(2\dot{H}+3H^2)=K+p_m,
\end{equation}
\begin{equation}
(K_X+2XK_{XX})  \dot{X}+6HXK_X-K_{\phi}=0,
\end{equation}
\begin{equation}
    \dot{\rho_m}+3H(\rho_m+p_m)=0,
\end{equation}
where $X=-(1/2)\dot{\phi}^2$. The equation for the scalar field $\phi$ is given by
\begin{equation}
    -(a^{3}\dot{\phi}K_{X})_{t}=a^{3}K_{\phi},
\end{equation}
which corresponds to the equation (3.4). In the pure kinetic k-essence case we have $K_{\phi}=0$ and from the last equation one has (see, e.g. \cite{L1})
\begin{equation}
a^{3}\dot{\phi}K_{X}=a^{3}\sqrt{-2X}K_{X}=\sqrt{\kappa}=const.
\end{equation}

\section{Models of k-essence for FRW universe}

In what follows we will present some new models of k-essence. All of they may give rise to cosmic acceleration.

\subsection{Example 1: The M$_{12}$ - model}
Let us consider  the   M$_{12}$ - model with the  following Lagrangian
  \begin {equation}
K=\nu_{-m}(N)N^{-m}+ ... +\nu_{-1}(N)N^{-1}+\nu_0(N)+\nu_1(N)N+ ... + \nu_n(N)N^n,
\end{equation}
where  in general $\nu_j=\nu_j(\phi)=\nu_j(N)$ and $N=\ln{(aa_0^{-1})}$. We study the case $m=0, n=2, \nu_j=const$. The M$_{12}$ - model becomes
\begin{align}
K=\nu_0+\nu_1 N+\nu_2 N^2.
\end{align}
To find $\nu_j$ and $X$ we look for  $H$ in the form
\begin{align}
    H=\mu_0+\mu_1N,
\end{align}
where $\mu_j=\textbf{consts}$ [in general $\mu_j=\mu_j(t)$]. This solution corresponds to the scale factor
\begin{align}
a=a_0e^{N}.
\end{align}
Finally,  we obtain the following parametric form of the M$_{12}$ - model (parametric pure kinetic k-essence)
\begin{align}
    K=-(2\mu_0\mu_1+3\mu_0^2)-2\mu_1(\mu_1+3\mu_0)N-3\mu_1^2N^2,
\end{align}
\begin{align}
X=k^{-1} a^{6}_0\mu_1^2(\mu_0+\mu_1 N)^2e^{6N}.
\end{align}

\subsection{Example 2: The M$_{1}$ - model}
Our next example is   the   M$_{1}$ - model, whose  Lagrangian assumes the form
  \begin {equation}
K=\nu_{-m}(t)t^{-m}+...+\nu_{-1}(t)t^{-1}+\nu_0(t)+\nu_1(t)t+ ... +\nu_n(t)t^n,
\end{equation}
where  in general $\nu_j=\nu_j(\phi)=\nu_j(t)$.
Let us explore  this model for  the case: $m=0, n=2$ and $\nu_j=consts.$ In this case  the M$_1$ - model takes the form
\begin{align}
K=\nu_0+\nu_1 t+\nu_2 t^2.
\end{align}
To find $\nu_j$ and $X$ we look for the following form of $H$,
\begin{align}
    H=\mu_0+\mu_1t
\end{align}
so that
\begin{align}
a=a_0e^{\mu_0t+0.5\mu_1 t^2},
\end{align}
where $\mu_j=consts$ [in general $\mu_j=\mu_j(t)$].
As a consequence, we obtain the following explicit form of the k-essence Lagrangian
\begin{align}
    K=-(2\mu_1+3\mu_0^2)-6\mu_0\mu_1t-3\mu_1^2t^2.
\end{align}
We also have
\begin{align}
2XK_X=3H^2+K=-2\dot{H}=-2\mu_1.
\end{align}
For $X$ we get the following expression
\begin{align}
X=\gamma_2^{-1}e^{6\mu_0t+3\mu_1t^2}, \quad \gamma_2^{-1}=\kappa^{-1}a_0^6\mu_1^2\,,
\end{align}
from which
\begin{align}
t=\frac{1}{3\mu_1}[-3\mu_0\pm\sqrt{9\mu_0^2+3\mu_1\ln{(\gamma_2X)}}].
\end{align}
Finally, we reconstruct the M$_{23}$ - model
\begin{align}
    K=-2\mu_1-3\mu_0^2-\mu_1\ln[\gamma_2X]=\nu_0+\nu_1\ln{X}.
\end{align}
[We recall that in general the M$_{23}$-model is read as
  \begin {equation}
K=\nu_{-m}(t)\zeta^{-m}+...+\nu_{-1}(t)\zeta^{-1}+\nu_0(t)+\nu_1(t)\zeta+ ... +\nu_n(t)\zeta^n,
\end{equation}
where $\zeta=\ln{X}$.]

\subsection{Example 3: The M$_{24}$ - model}
Here we present  the  M$_{24}$ - model
  \begin {equation}
K=\frac{2m\lambda\sigma^2(-2\beta v+\lambda v^2+\lambda)(1-v^2)}{(\beta-\lambda v)^2}-3[n-\frac{m\lambda\sigma(1-v^2)}{\beta-\lambda v}]^2,
\end{equation}
\begin{align}
X=\gamma_3(2\beta v-\lambda v^2-\lambda)^2(1-v^2)^2(\beta-\lambda v)^{6m-4},
\end{align}
where $ \gamma_3=\kappa^{-1}\alpha^6m^2\lambda^2\sigma^6$,  $v=\tanh[\sigma t]$ and $\lambda, \sigma, \alpha, \beta, n, m$ are some constants.  Solving the equation (3.3) we obtain
\begin{align}
    H=n-\frac{m\lambda\sigma(1-v^2)}{\beta-\lambda v}
\end{align}
from which we derive the scale factor as
\begin{align}
a=\alpha[\beta-\lambda v]^me^{nt}.
\end{align} Note that
\begin{align}
\dot{H}=\frac{m\lambda\sigma^2(2\beta v-\lambda v^2-\lambda)(1-v^2)}{(\beta-\lambda v)^2}.
\end{align}

\section{The relation  between $F(T)$-gravity and k-essence in the FRW universe}

 In this section, we want to analyze the relation between modified teleparallel gravity and pure kinetic k-essence. In Appendix C, we will consider this relation in the context of general modified gravity theories.
\subsection{General case}
\subsubsection{Version-I}
Let us consider the following transformation
\begin {equation}
K=8\dot{H}Tf_{TT}-2(T-2\dot{H})f_{T}+f,
\end{equation}
\begin {equation}
X=\kappa^{-1}k^{-4}a^{6}[\dot{H}+0.5k^2(\rho_m+p_m)]^2,
\end{equation}
where $ T=-6H^2.$ Thus Eqs.(2.12)-(2.14) take the form
\begin {equation}
0=-3\kappa^{-2}H^{2}+2XK_X-K+\rho_{m},
\end{equation}
\begin {equation}
0=\kappa^{-2}(2\dot{H}+3H^2)+K+p_m,
\end{equation}
\begin{equation}
(K_X+2XK_{XX})  \dot{X}+6HXK_X=0,
\end{equation}
\begin{equation}
    \dot{\rho_m}+3H(\rho_m+p_m)=0.
\end{equation}
These  are  the equations of motion of pure kinetic k-essence. This result shows that the field equations of modified teleparallel gravity and pure kinetic k-essence are equivalent to each other. This equivalence permits to construct a new class of pure kinetic k-essence models starting from some models of  modified teleparallel gravity.
Let us see it  for the following modified teleparallel gravity model: $f(T)=\alpha T^n$ \cite{BF}-\cite{L3}. In this case, we have
\begin{align}
    f_T=\alpha n T^{n-1}, \quad f_{TT}=\alpha n(n-1)T^{n-2}.
\end{align}
Substituting these expressions into the equations (4.1)-(4.2) we get
\begin {equation}
K=8\alpha n(n-1)\dot{H}T^{n-1}-2\alpha n(T-2\dot{H})T^{n-1}+\alpha T^n,
\end{equation}
\begin {equation}
X=\kappa^{-1}k^{-4}a^{6}[\dot{H}+0.5k^2(\rho_m+p_m)]^2.
\end{equation}
Let us consider some specific case.\\
i) If the scale factor behaves as $a=a_0e^{g(t)}$ so that $H=\dot{g}, \dot{H}=\ddot{g}$,  $K$ and $X$ take the form
\begin {equation}
K=8\alpha n(n-1)\ddot{g}(-6)^{n-1}\dot{g}^{2(n-1)}-2\alpha n(-6\dot{g}^2-2\ddot{g})(-6)^{n-1}\dot{g}^{2(n-1)}+\alpha (-6)^n\dot{g}^{2n},
\end{equation}
\begin {equation}
X=\kappa^{-1}k^{-4}a^{6}\ddot{g}^2.
\end{equation}
If we now consider the simplest case $g=t$ (it means, $\dot{g}=1, \ddot{g}=0$),  we get
\begin {equation}
K=-2\alpha n(-6)^{n}+\alpha (-6)^n=(1-2 n)\alpha (-6)^n,
\end{equation}
\begin {equation}
X=0.
\end{equation}

ii) A non-trivial model may be obtained from $a=a_0t^m$. In this case $H=mt^{-1}, \dot{H}=-mt^{-2}, T=\frac{-6m^2}{t^2}$ and  $K$ and $X$ take the form
\begin {equation}
K=8\alpha n(n-1)\dot{H}(\frac{-6m^2}{t^2})^{n-1}-2\alpha n(\frac{-6m^2}{t^2}-2\dot{H})(\frac{-6m^2}{t^2})^{n-1}+\alpha (\frac{-6m^2}{t^2})^n,
\end{equation}
\begin {equation}
X=\kappa^{-1}k^{-4}a_0^{6}m^2t^{6m-4}
\end{equation}
or
\begin {equation}
K=2\alpha m(-6m^2)^{n-1}[-4n(n-1)+2n(1-3m)+3m]t^{-2n},
\end{equation}
\begin {equation}
X=\kappa^{-1}k^{-4}a_0^{6}m^2t^{6m-4}=\gamma_5^{-1}t^{6m-4}.
\end{equation}
Since $t=(\gamma_5 X)^{\frac{1}{6m-4}}$ we finally get the following pure kinetic k-essence model
\begin {equation}
K=2\alpha m(-6m^2)^{n-1}[-4n(n-1)+2n(1-3m)+3m](\gamma_5 X)^{\frac{n}{2-3m}}.
\end{equation}
\subsubsection{Version-II}
Let us rewrite Eqs.(2.12)-(2.14) as
\begin {equation}
3k^{-2}H^{2}=\rho_{eff}+\rho_{m},
\end{equation}
\begin {equation}
-\kappa^{-2}(2\dot{H}+3H^2=p_{eff}+p_m,
\end{equation}
\begin{equation}
    \dot{\rho}_m+3H(\rho_m+p_m)=0,
\end{equation}
where
\begin {equation}
\rho_{eff}=2T f_{T}-f, \quad p_{eff}=8\dot{H}Tf_{TT}-2(T-2\dot{H})f_{T}+f.
\end{equation}
We introduce the following two functions $K$ and $X$,
\begin {equation}
K=8\dot{H}Tf_{TT}-2(T-2\dot{H})f_{T}+f, \quad X=\frac{4\dot{H}^2(2Tf_{TT}+f_{T})^2}{\kappa a^{-6}}.
\end{equation}
These functions belongs to the system of the equations (4.3)-(4.6).
\subsection{Specific case  case: $\phi=\phi_0+\ln{a^{\pm\sqrt{12}}}$}
One specific interesting  case is given by
\begin{align}
    \phi=\phi_0+\ln{a^{\pm\sqrt{12}}}.
\end{align}
 It deserves  separate investigation. In fact for this case $\dot{\phi}=\pm\sqrt{12}H$ so that $X=-0.5\dot{\phi}^2=-6H^2=T$.
 The  corresponding continuity equation is
\begin{align}
    \ddot{\phi}(f_T-\dot{\phi}^2f_{TT})+3H\dot{\phi}f_T=0
\end{align}
or, in terms of $T$,
\begin{align}
    (f_T+2Tf_{TT})\dot{T}+6HTf_T=0,
\end{align}
where $\rho^{\prime}=2Tf_T-f, \quad p^{\prime}=f$ and $\dot{\rho}^{\prime}+3H(\rho^{\prime}+p^{\prime})=0$. Let us split the equation (2.13) into two separate equations,
\begin{align}
    4\dot{H}Tf_{TT}-(T-2\dot{H})f_{T}=0
\end{align}
and
\begin{align}
    -4\dot{H}+T-f=2\kappa^2p_m.
\end{align}
Eq.(4.27) is automatically satisfied since it is just an another form for the continuity equation (4.26). So we finally obtain the equation system for $F(T)$ - gravity, which
takes the form
\begin{align}
    -T-2Tf_{T}+f=2\kappa^2 \rho_m,
\end{align}
\begin{align}
    -4\dot{H}+T-f=2\kappa^2p_m,
\end{align}
\begin{align}
    (f_T+2Tf_{TT})\dot{T}+6HTf_T=0,
\end{align}
\begin{align}
    \dot{\rho}_m+3H(\rho_m+p_m)=0.
\end{align}
After the identification $T=X=-6H^2$ and  $f=2\kappa^{2}K$, we recover equations (4.3)-(4.6) . So we can conclude that for the special case (4.24) both  $F(T)$ - gravity and pure kinetic k-essence are equivalent to  each  other at least at the level of the dynamical
 equations. Some remarks can be observed from the continuity equation (4.25) [=(4.26)=(4.27)]. Two integrals of motion ($I_{jT}=0$) appear:
\begin{align}
I_1=a_0^{-3}a^3T^{0.5}f_T, \quad I_2=   f-a^3T^{0.5}f_T\partial_T^{-1}(a^{-3}T^{-0.5}).
\end{align}
Their general solution   is given by
\begin{align}
f=C_2+iC_1a_0^2\partial_T^{-1}(a^{-3}T^{-0.5}), \quad C_j=const.
\end{align}
Finally we would like to present an exact solution for both $F(T)$-gravity and pure kinetic k-essence.  Let us consider  the $\Lambda$CDM model for which $a^{-3}=-\frac{1}{2\rho_0}(T+2\Lambda)=-\frac{1}{2\rho_0}(X+2\Lambda)$ so that
\begin{align}
f=f(X)=f(T)= C_2-\frac{iC_1a_0^{3}}{3\rho_0}(T^{1.5}+6\Lambda T^{0.5})= C_2-\frac{iC_1a_0^{3}}{3\rho_0}(X^{1.5}+6\Lambda X^{0.5}),
\end{align}
which is the M$_{32}$ - model. This is the exact solution of the equations of motion of pure kinetic k-essence and $F(T)$ - gravity simultaneously.

\section{  $F(R,T)$ gravity}

We have just considered one generalization of $F(T)$ in the presence of scalar field. In this section we would like to present another possible generalization of $F(T)$ gravity, namely the so-called $F(R,T)$ gravity.
\subsection{The  M$_{37}$ - model}
 The  action of M$_{37}$ - gravity is given by \cite{M7}
\begin{equation}\label{2.1}
 S_{37}=\int d^4 x\sqrt{-g}[F(R,T)+\mathcal{L}_m],
 \end{equation}
 where $\mathcal{L}_m$ is the matter Lagrangian, $\epsilon_i=\pm 1$ (signature) and
 \begin{eqnarray}
 R&=&u+\epsilon_1g^{\mu\nu}R_{\mu\nu},\label{2.2}\\
   T&=&v+\epsilon_2{S_\rho}^{\mu\nu}\,{T^\rho}_{\mu\nu},\label{2.3}
 \end{eqnarray}
Here $u=u(x_i; g_{ij}, \dot{g_{ij}},\ddot{g_{ij}}, ... ; f_j)$ and  $v=v(x_i; g_{ij}, \dot{g_{ij}},\ddot{g_{ij}}, ... ; g_j)$  are some functions to be defined.
 Now we work in the FRW universe with the metric (2.9).
  In this case the curvature and torsion scalars can be written as
   \begin{eqnarray}
 R&=&u+6\epsilon_1(\dot{H}+2H^2),\label{2.5}\\
   T&=&v+6\epsilon_2H^2,\label{2.6}
 \end{eqnarray}
where, $u=u(t, a,\dot{a}, \ddot{a},\dddot{a}, ...; f_i)$ and $v=v(t,a,\dot{a}, \ddot{a},\dddot{a}, ...; g_i)$ are some real functions, $H=(\ln a)_t$,  while $f_i$ and $g_i$ are some unknown functions related with the geometry of the spacetime. By introducing the Lagrangian multipliers we can now rewrite the action (\ref{2.1}) as
 \begin{equation}\label{2.7}
 S_{37}=\int dt\,a^3\left\{F(R,T)-\lambda\left[T-v-
 6\epsilon_2\frac{\dot{a}^2}{a^2}\right]-\gamma\left[R-u- 6\epsilon_1\left(\frac{\ddot{a}}{a}+\frac{\dot{a}^2}{a^2}\right)\right]+L_m\right\},
 \end{equation}
 where $\lambda$ and $\gamma$ are Lagrange multipliers. If we take the variations with
 respect to $T$ and $R$ of this action we get
  \begin{equation}\label{2.8}
 \lambda=F_T, \quad \gamma=F_R.
 \end{equation}
 Therefore, the action~(\ref{2.7}) can be rewritten as
 \begin{equation}
 S_{37}=\int dt\,a^3\left\{F(R,T)-F_T\left[T-v-
 6\epsilon_2\frac{\dot{a}^2}{a^2}\right]-F_R\left[R-u- 6\epsilon_1\left(\frac{\ddot{a}}{a}+\frac{\dot{a}^2}{a^2}\right)\right]+L_m\right\}.
\end{equation}
 Then the  corresponding  point-like Lagrangian reads as
 \begin{equation}
 L_{37}=a^3[F-(T-v)F_T-(R-u)F_R+L_m]-
 6(\epsilon_1F_R-\epsilon_2F_T) a\dot{a}^2-6\epsilon_1(F_{RR}\dot{R}+F_{RT}\dot{T})a^2\dot{a}.
\end{equation}
We finally obtain the following equations of the M$_{37}$ -  model \cite{M7}:
\begin{eqnarray}
 D_2F_{RR}+D_1F_R+JF_{RT}+E_1F_T+KF&=&-2a^3\rho,\nonumber\\
   U+B_2F_{TT}+B_1F_{T}+C_2F_{RRT}+C_1F_{RTT}+C_0F_{RT}+MF  &=&6a^2p,\label{1.12}\\
 \dot{\rho}+3H(\rho+p)&=&0.\nonumber
 \end{eqnarray}
 Here
 \begin{eqnarray}
 D_2&=&-6\epsilon_1\dot{R}a^2\dot{a},\\
   D_1&=&6\epsilon_1a^2 \ddot{a}+a^3u_{\dot{a}}\dot{a},\\
  J&=&-6\epsilon_1a^2 \dot{a}\dot{T},\\
 E_1&=&12\epsilon_2a \dot{a}^2+a^3v_{\dot{a}}\dot{a},\\
 K&=&-a^3
 \end{eqnarray}
and
\begin{eqnarray}
 U&=&A_3F_{RRR}+A_2F_{RR}+A_1F_{R},\\
 A_3&=&-6\epsilon_1\dot{R}^2a^2,\label{eq17}\\
 A_2&=&-12\epsilon_1\dot{R}a\dot{a}-6\epsilon_1\ddot{R}a^2+a^3\dot{R}u_{\dot{a}},\\
   A_1&=&12\epsilon_1\dot{a}^2+6\epsilon_1a \ddot{a}+3a^2\dot{a}u_{\dot{a}}+a^3\dot{u}_{\dot{a}}-a^3u_a,\\
 B_2&=&12\epsilon_2\dot{T}a \dot{a}+a^3\dot{T}v_{\dot{a}},\\
 B_1&=&24\epsilon_2\dot{a}^2+12\epsilon_2a \ddot{a}+3a^2\dot{a}v_{\dot{a}}+a^3\dot{v}_{\dot{a}}-a^3v_a,\\
 C_2&=&-12\epsilon_1a^2\dot{R}\dot{T},\\
 C_1&=&-6\epsilon_1a^2\dot{T}^2,\\
 C_0&=&-12\epsilon_1\dot{T}a\dot{a}+12\epsilon_2\dot{R}a\dot{a}-6\epsilon_1a^2\ddot{T}+a^3\dot{R}v_{\dot{a}}+a^3\dot{T}u_{\dot{a}},\\
 M&=&-3a^2.
 \end{eqnarray}
 The M$_{37}$ - model (\ref{2.1}) admits some interesting particular and physically important  cases. Let us see some example.

 i) \textit{$F(R)$ - gravity}.
 If the model is independent of the torsion, namely $F=F(R,T)=F(R)$, and we assume that $u=0$, the  action (\ref{2.1}) takes the form
   \begin{equation}\label{3.1}
 {\cal S}_{R}=\int d^4 x e[F(R)+L_m],
 \end{equation}
 where
 \begin{equation}\label{3.2}
 R=\epsilon_1g^{\mu\nu}R_{\mu\nu},
 \end{equation}
  is the curvature scalar. We work with  the FRW metric \eqref{2.4}. In this case  $R$ assumes the form
 \begin{equation} \label{3.3}
R=6(\dot{H}+2H^2).
\end{equation}
We rewrite the action as
 \begin{equation}\label{3.4}
 {\cal S}_R=\int dtL_R,
 \end{equation}
 where the   Lagrangian is given by
 \begin{equation}\label{3.5}
  L_R=a^3(F-RF_R+L_m)-
 6\epsilon_1F_R a\dot{a}^2-6\epsilon_1F_{RR}\dot{R}a^2\dot{a}.
 \end{equation}
 The corresponding field equations of $F(R)$ gravity read
 \begin{eqnarray}
 6\dot{R}HF_{RR}-(R-6H^2)F_R+F&=&\rho,\label{2.5}\\
   -2\dot{R}^2F_{RRR}+[-4\dot{R}H-2\ddot{R}]F_{RR}+[-2H^2-4a^{-1} \ddot{a}+R]F_{R}-F &=&p,\label{2.6}\\
 \dot{\rho}+3H(\rho+p)&=&0.\label{2.77}
 \end{eqnarray}

 ii) \textit{$F(T)$ - gravity}.
Now we assume that the function $F=F(R,T)$ is independent of the curvature scalar $R$ and $v=0$. In this case we get the modified teleparallel gravity - $F(T)$ gravity. Its  gravitational action is
 \begin{equation}\label{2.8}
 {\cal S}_T=\int d^4 xe [F(T)+L_m],
 \end{equation}
 where $e={\rm det}\,(e_\mu^i)=\sqrt{-g}$.
 The torsion scalar $T$ is defined as
 \begin{equation}\label{2.9}
 T=\epsilon_2{S_\rho}^{\mu\nu}\,{T^\rho}_{\mu\nu}\,,
 \end{equation}
 where
 \begin{eqnarray}
 {T^\rho}_{\mu\nu} &\equiv &-e^\rho_i\left(\partial_\mu e^i_\nu
 -\partial_\nu e^i_\mu\right)\,,\label{2.10}\\
 {K^{\mu\nu}}_\rho &\equiv &-\frac{1}{2}\left({T^{\mu\nu}}_\rho
 -{T^{\nu\mu}}_\rho-{T_\rho}^{\mu\nu}\right)\,,\label{eq6}\label{2.11}\\
 {S_\rho}^{\mu\nu} &\equiv &\frac{1}{2}\left({K^{\mu\nu}}_\rho
 +\delta^\mu_\rho {T^{\theta\nu}}_\theta-
 \delta^\nu_\rho {T^{\theta\mu}}_\theta\right)\,.\label{2.12}
 \end{eqnarray}
 For a spatially flat FRW metric \eqref{2.4},  we have that the torsion scalar  assumes  the form
 \begin{equation}\label{2.13}
 T=T_s=-6H^2.
 \end{equation}
 The action \eqref{2.8} can be written as
 \begin{equation}\label{2.14}
 {\cal S}_T=\int dt L_T,
 \end{equation}
where
 the point-like Lagrangian reads
 \begin{equation}\label{2.15}
  L_T=a^3\left(F-F_T T\right)-
 6F_T a\dot{a}^2-a^3L_m.
 \end{equation}
  The equations of F(T) gravity look like
 \begin{eqnarray}
 12H^2 F_T+F&=&\rho,\label{2.16}\\
 48H^2 F_{TT}\dot{H}-F_T\left(12H^2+4\dot{H}\right)-F
 &=&p,\label{2.17}\\
 \dot{\rho}+3H(\rho+p)&=&0.\label{2.18}
 \end{eqnarray}

 \subsection{The M$_{43}$ - model}
In this subsection we consider the M$_{43}$ - model which  is one of the  representatives of $F(R,T)$ gravity.
 The  action of M$_{43}$ - model reads as
\begin{eqnarray}
 S_{43}&=&\int d^4 x\sqrt{-g}[F(R,T)+L_m],\nonumber\\
 R&=&R_s=\epsilon_1g^{\mu\nu}R_{\mu\nu},\label{2.1}\\
   T&=&T_s=\epsilon_2{S_\rho}^{\mu\nu}\,{T^\rho}_{\mu\nu},\nonumber
 \end{eqnarray} where $L_m$ is the matter Lagrangian, $\epsilon_i=\pm 1$ (signature), $R$ is the curvature scalar, $T$ is the torsion scalar. Let us  consider
the spacetime where the curvature and torsion are written by using
the connection  $G^\lambda {}_{\mu \nu }$ as a sum of  the curvature and
torsion, namely
\begin{equation}\label{2.4}
G^\lambda {}_{\mu \nu }=e_i{}^\lambda \partial _\mu
e^i{}_\nu
+e_j{}^\lambda e^i{}_\nu \omega{}^j{}_{i\mu }=\Gamma^\lambda {}_{\mu \nu }+K^\lambda {}_{\mu \nu}.
 \end{equation}
Here $\Gamma^{j}_{i\mu}$ is the Levi-Civita connection and $K^{j}_{i\mu}$ is the contorsion.
 The quantities $\Gamma^{j}_{i\mu}$ and  $K^{j}_{i\mu}$ are defined as
 \begin{equation}\label{2.9}
\Gamma^l{}_{jk}=\frac{1}{2}g^{lr} \{\partial _k g_{rj} + \partial _j
g_{rk} - \partial _r g_{jk} \},
\end{equation}
 and
 \begin{equation}\label{2.10}
K^{\lambda}_{\mu\nu}=-\frac 12\left( T^\lambda {}_{\mu \nu
}+T_{\mu
\nu }{}^\lambda +T_{\nu \mu }{}^\lambda \right),
\end{equation}
respectively.
Here the components of the torsion tensor are given by
 \begin{eqnarray}\label{2.11}
T^\lambda {}_{\mu \nu } &=&e_i{}^\lambda T^i{}_{\mu \nu }=\Gamma
^\lambda
{}_{\mu \nu }-\Gamma ^\lambda {}_{\nu \mu },   \\ \label{2.12}
T{}^i{}_{\mu \nu } &=&\partial _\mu e{}^i{}_\nu -\partial _\nu
e{}^i{}_\mu +\Gamma {}^i{}_{j\mu }e{}^j{}_\nu -\Gamma {}^i{}_{j\nu
}e{}^j{}_\mu .
\end{eqnarray}

The curvature $R^\rho {}_{\sigma \mu \nu }$ is defined  as
\begin{eqnarray} \label{2.13}
R^\rho {}_{\sigma \mu \nu } &=&e_i{}^\rho e^j{}_\sigma R^i{}_{j\mu
\nu }=\partial _\mu G^\rho {}_{\sigma \nu }-\partial _\nu
G^\rho {}_{\sigma \mu }+G^\rho {}_{\lambda \mu }G^\lambda {}_{\sigma \nu }-G^\rho {}_{\lambda \nu }G^\lambda {}_{\sigma \mu }
\nonumber \\
&=&\bar{R}^\rho {}_{\sigma \mu \nu }+\partial _\mu
K^\rho {}_{\sigma \nu }-\partial _\nu K^\rho {}_{\sigma \mu }+K^\rho
{}_{\lambda \mu }K^\lambda {}_{\sigma \nu }-K^\rho {}_{\lambda \nu
}K^\lambda {}_{\sigma
\mu }  \nonumber \\
&&+\Gamma^{\rho}_{\lambda\mu} K^\lambda {}_{\sigma \nu
}-\Gamma^{\rho}_{\lambda\nu} K^\lambda {}_{\sigma \mu
}+\Gamma^{\lambda}_{\sigma\nu} K^\rho {}_{\lambda \mu
}-\Gamma^{\lambda}_{\sigma\mu} K^\rho {}_{\lambda \nu
},
\end{eqnarray}
where  the Riemann curvature is
defined in the standard way
\begin{equation} \label{2.14}
\bar{R}^\rho {}_{\sigma \mu \nu }=\partial
_\mu \Gamma^{\rho}_{\sigma\nu} -\partial _\nu \Gamma^{\rho}_{\sigma\mu} +\Gamma^{\rho}_{\lambda\mu}
\Gamma^{\lambda}_{\sigma\nu} -\Gamma^{\rho}_{\lambda\nu} \Gamma^{\lambda}_{\sigma\mu}. \end{equation}
 Now we introduce the curvature and torsion scalars,
\begin{eqnarray} \label{2.15}
R&=&g^{ij}R_{ij}, \\   \label{2.16}
T&=&S_\rho^{\mu\nu}T^\rho_{\mu\nu},
\end{eqnarray}
where
\begin{equation} \label{2.17}
        S_\rho^{\mu\nu}=\frac{1}{2}\left(K^{\mu\nu}_\rho+\delta_{\rho}^\mu T^{\theta\nu}_\theta-\delta_{\rho}^\nu T^{\theta\mu}_\theta\right).
\end{equation}
Now the M$_{43}$ - model is written in the form of \eqref{2.1}.

Now we want to present the M$_{43}$ - model for the spatially flat FRW spacetime. In this case the metric assumes the form
\begin{equation}\label{2.18}
 ds^2=-dt^2+a^2(t)(dx^2+dy^2+dz^2),
 \end{equation}where $a(t)$ is the scale factor. In this case,
 the non-vanishing components of the Levi-Civita connection are
\begin{eqnarray}\label{2.19}
\Gamma^{0}_{00}&=&\Gamma^{0}_{0i}=\Gamma^{0}_{i0}=\Gamma^{i}_{00}=\Gamma^{i}_{jk}=0,\nonumber \\
\Gamma^{0}_{ij}&=& a^2H
\delta _{ij},  \\
\Gamma^{i}_{jo}&=&\Gamma^{i}_{0j}=H\delta _j^i,\nonumber
\end{eqnarray}
where $H=(\ln a)_t$ and $i,j,k,...=1,2,3.$ Let us calculate the
components of torsion tensor. The non-vanishing  components  are
given by:
\begin{eqnarray} \label{2.20}
T_{110} &=&T_{220}=T_{330}=a^2h, \nonumber  \\
T_{123} &=&T_{231}=T_{312}=2a^3f,
\end{eqnarray}
where  $h$ and  $f$ are some real functions.
 Note that  the indices of the torsion tensor are raised and lowered with  the metric, namely
\begin{equation}\label{2.21}
 T_{ijk}=g_{kl}T_{ij}{}{}^{l}.
 \end{equation}

Now we can find  the contortion components. We get
\begin{eqnarray} \label{2.22}
K^1{}_{10} &=&K^2{}_{20}=K^3{}_{30}=0,  \nonumber \\
K^1{}_{01} &=&K^2{}_{02}=K^3{}_{03}=h,  \nonumber \\
K^0{}_{11} &=&K^0{}_{22}=K^0{}_{22}={}a^2h, \\
K^1{}_{23} &=&K^2{}_{31}=K^3{}_{12}=-af,  \nonumber \\
K^1{}_{32} &=&K^2{}_{13}=K^3{}_{21}=af. \nonumber
\end{eqnarray}

The non-vanishing components of the curvature $R^\rho {}_{\sigma \mu
\nu }$  are given by
\begin{eqnarray} \label{2.23}
R^0{}_{101} &=&R^0{}_{202}=R^0{}_{303}=a^2( \dot{H}%
+H^2+Hh+\dot{h}),  \nonumber \\
R^0{}_{123} &=&-R^0{}_{213}=R^0{}_{312}=2a^3f( H+h),
\nonumber
\\
R^1{}_{203} &=&-R^1{}_{302}=R^2{}_{301}=-a( Hf+\dot{f}
),  \nonumber \\
R^1{}_{212} &=&R^1{}_{313}=R^2{}_{323}=a^2[ ( H+h)
^2-f^2].
\end{eqnarray}
Similarly, we write the non-vanishing components of the  Ricci
curvature tensor $R{}_{\mu \nu }$ as
\begin{eqnarray} \label{2.24}
R{}_{00} &=&-3\dot{H}-3\dot{h}-3H^2-3Hh,
\nonumber
\\
R{}_{11} &=&R{}_{22}=R{}_{33}=a^2( \dot{H}+\dot{h}+3H^2+5Hh+2h^2-f^2).
\end{eqnarray}
The non-vanishing components of the tensor $S_\rho^{\mu\nu}$ are
\begin{eqnarray} \label{2.25}
S_1^{10}&=&\frac{1}{2}\left(K^{10}_1+\delta_1^1T^{\theta 0}_\theta-\delta_1^0T^{\theta\nu}_\theta\right)=\frac{1}{2}\left(h+2h\right)=h,\\ \label{2.26}
S_1^{10}&=&S_2^{20}=S_3{^30}=2h,\\ \label{2.27}
S_1^{23}&=&\frac{1}{2}\left(K^{23}_1+\delta_1^2+\delta_1^3\right)=-\frac{f}{2a},\\ \label{2.28}
S_1^{23}&=&S_2^{31}=S_3^{21}=-\frac{f}{2a}
\end{eqnarray}
and
\begin{eqnarray}\label{2.29}
T=T^1_{10}S_1^{10}+T^2_{20}S_2^{20}+T^3_{30}S_3^{30}+T_1^{23}S^1_{23}+T^2_{31}S_2^{31}+T^3_{12}S_3^{12}.
\end{eqnarray}
Now we can write the explicit forms of the curvature and torsion scalars. One has
\begin{eqnarray} \label{2.30}
R&=&6(\dot{H}+2H^2) +6\dot{h}+18Hh+6h^2-3f^2 \\ \label{2.31}
T&=&6h^2-a^{-2}f^2.
\end{eqnarray}
For FRW metric, the M$_{43}$ - model  takes the form
\begin{eqnarray}
S_{43}&=&\int d^4 x\sqrt{-g}[F(R,T)+L_m], \nonumber\\
R&=&6(\dot{H}+2H^2) +6\dot{h}+18Hh+6h^2-3f^2, \label{2.32}\\
T&=&6(h^2-f^2).\nonumber
\end{eqnarray}
In this way, we have derived the M$_{43}$ - model as  one of geometrical realizations of $F(R,T)$ gravity by starting from the pure geometrical point of view.

\section{Conclusion}
In this work we have presented a brief review on $F(T)$ gravity.  We
have investigated generalized $F(T)$ modified torsion models, that
is, models in which the torsion gravity  equations are extended to
scalar fields.  This study is a continuation of our investigation
program of $F(T)$  gravity \cite{M1}. We note that the GR case
corresponds not only to the model $F(T)=T$, but also to our specific
model $F(T)=\alpha T+\beta T^{1/2}$, for which we obtain the same
results.

We also considered the recently developed $F(T)$ gravity,  which is
a new modified gravity capable of accounting for the present cosmic
accelerating expansion.  In particular, we presented some new models
of $F(T)$ gravity and k-essence. We analyzed the relation between
$F(T)$ gravity and k-essence. We also  studied some new parametric
models of pure kinetic k-essence. We also presented a short review
on Noether symmetry of $F(T)$ gravity.  Finally we have considered
some generalizations of $F(T)$ gravity.

\end{document}